\documentclass[10pt,superscriptaddress,twocolumn,amsmath,amssymb,aps,prl,showpacs]{revtex4-1}
\usepackage{mathrsfs}
\usepackage{array}
\usepackage{graphicx}
\usepackage{dcolumn}
\usepackage{bm}
\usepackage[title,titletoc,header]{appendix}
\usepackage{amssymb}
\usepackage{amsmath}
\usepackage{mathrsfs}
\usepackage{amsmath}
\usepackage{subfigure}
\usepackage{float}
\usepackage[title,titletoc,header]{appendix}
\usepackage{graphicx}
\usepackage{color}
\usepackage{amssymb}
\usepackage{graphicx}
\usepackage{verbatim}

\begin{document}

\title{Effective theory and universal relations for Fermi gases near a $d$-wave  interaction  resonance
}
\author{Pengfei Zhang}
\affiliation{Institute for Advanced Study, Tsinghua University, Beijing, 100084, China}
\author{Shizhong Zhang}
\email[]{shizhong@hku.hk}
\affiliation{Department of Physics and Center of Theoretical and Computational Physics,
The University of Hong Kong, Hong Kong, China}
\author{Zhenhua Yu}
\email[]{huazhenyu2000@gmail.com}
\affiliation{Institute for Advanced Study, Tsinghua University, Beijing, 100084, China}
\date{\today }

\begin{abstract}
In this work, we present an effective field theory to describe a two-component Fermi gas near a $d$-wave interaction resonance. The effective field theory is renormalizable by matching with the low energy $d$-wave scattering phase shift. Based on the effective field theory, we derive universal properties of the Fermi gas by the operator product expansion method. 
We find that beyond the contacts defined by adiabatic theorems, the asymptotic expressions of the momentum distribution and the Raman spectroscopy involve two extra contacts which provide additional information of correlations of the system. Our formalism sets the stage for further explorations of many-body effects in a $d$-wave resonant Fermi gas.  Finally we generalize our effective field theory for interaction resonances of arbitrary higher partial waves. 
\end{abstract}

\maketitle

\emph{Introduction.} 
Correlations of $d$-wave symmetry are of fundamental interest in modern physics. One outstanding example is the $d$-wave Cooper pairing in high-$T_c$ superconductors which provides a paradigmatic case of strongly correlated electron systems~\cite{HTC}. In cold atom systems, strong $d$-wave correlations can also be generated close to a $d$-wave Feshbach resonance, as has been demonstrated experimentally in Cr~\cite{Pfau, Gorceix}. While it is generally believed that, compared with $s$-wave resonance, atomic gases close to higher partial wave resonances suffer more rapid atom loss, recent spectroscopic measurements around a $p$-wave Feshbach resonance indicate that quasi-equilibrium states of such systems exist and their universal properties can be investigated~\cite{Thywissen}. Theoretically, however, many-body physics with  resonant $d$-wave interactions  has been rarely studied and, in particular, an appropriate minimal model is still lacking.

In this work, we consider a two-component Fermi gas near a $d$-wave  interaction  resonance. We construct an effective low-energy field theory, the bare coupling constants of which are renormalized by matching with the $d$-wave scattering phase shift $\cot\delta(k)=-1/(Dk^5)-1/(vk^3)-1/(Rk)$. The super volume $D$, the effective volume $v$ and the effective range $R$ are the minimal set of parameters that is needed to parametrize the inter-fermion interactions. Furthermore, we use the effective theory, combined with the operator product expansion (OPE) method, to derive universal properties of the Fermi gas when the average inter-particle distance is much larger than the range $r_0$ associated with the inter-fermion interaction. We find that the universal behaviour of the system is  governed by five quantities, three of which are related to the variation of the system energy with respect to the three $d$-wave scattering parameters, analogous to the contacts defined in the case of $s$- and $p$-wave case~\cite{Tan2008,Braaten2008a, Zhang2009, Werner2009,Braaten2010, Valiente2011, Valiente2012, Ueda2015,Yu2015,Qi2016,Cui2016}. However,  we find that the sub-leading terms of the tails of momentum distribution and Raman spectroscopy involve two new contacts, which further characterise the correlations of the system at short distances.  Our effective field theory provides a minimal model for studying other many-body physics of Fermi gases near a $d$-wave resonance. We show that the $d$-wave contacts reveal much richer correlation structures than the $s$-wave case. Finally we generalize our formalism for resonant interactions to arbitrary higher partial waves. 

\emph{Effective field theory.} To describe the low energy degrees of freedom close to a $d$-wave interaction resonance, we adopt a Lagrangian field theory and requires that the Lagrangian density to obey the following symmetry requirements: (1) Rotation symmetry. (2) Galilean invariance such that the scattering of two fermions in vacuum does not depend on their center of mass momentum. In addition, we aim to establish a \emph{local} effective field theory, which should be \emph{renormalizable} in the low energy limit in terms of the minimal set of scattering parameters $D,v,R$, describing the $d$-wave scattering phase shift.

The Lagrangian density of the effective field theory that we construct for the system up to a momentum cutoff $\Lambda$ is given by 
\begin{align}
\mathcal{L}=
&\sum_{i=1}^{2}\psi_{i}^{\dagger}\left(i\partial_{t}+\dfrac{\nabla^{2}}{2M}\right)\psi_{i}+\sum_{m=-\ell}^{\ell}\bar g(d_{\ell m}^{\dagger}\mathcal Y_{m}+h.c.) \nonumber\\
+&\eta\sum_{m=-\ell}^{\ell} d_{\ell m}^{\dagger}\left[i\partial_{t}+\dfrac{\nabla^{2}}{4M}+\bar z\left(i\partial_{t}+\dfrac{\nabla^{2}}{4M}\right)^{2}-\bar\nu\right]d_{\ell m}
\label{L}
\end{align}
where $\ell=2$ and the operator $\mathcal Y_{m}$ is given by
\begin{align}
\mathcal Y_{m}=\dfrac{1}{4}\sum_{a,b=x,y,z}&C^{m}_{ab}[(\partial_{a}\psi_{1})(\partial_{b}\psi_{2})-(\partial_{a}\partial_{b}\psi_{1})\psi_{2}\notag \\
&+(\partial_{b}\psi_{1})(\partial_{a}\psi_{2})-\psi_{1}(\partial_{a}\partial_{b}\psi_{2})].
\label{Y}
\end{align}
The field operator $\psi_{i}$ is the annihilation operator for fermions in state $|i\rangle$. $M$ is the mass of the fermions. We take $\hbar=1$ throughout. The dimer fields $d_{\ell m}$ of azimuthal quantum number $m$ mediate the $d$-wave interaction between the two fermions, which we assume to be isotropic. $C^{m}_{ab}$ are the Clebsch-Gordon coefficients when transforming $k_{i}k_{j}/k^{2}$ to the spherical harmonics $\sqrt{4\pi}Y_{2m}(\hat{k})$. In terms of $a_{i,\mathbf k}$ and $b_{\ell m,\mathbf k}$, the Fourier transformations of the operators $\psi_{i}$ and $d_{\ell m}$, the fermion-dimer coupling in the Lagrangian $L=\int d{\bf r}\mathcal{L}$ [the second term in Eq.~(\ref{L})] takes the form 
\begin{align}
L_{fd}=\bar g\sqrt{\frac{4\pi}V}\sum_{m=-\ell}^\ell\sum_{\mathbf p,\mathbf k} [k^\ell Y_{\ell m}(\hat k)b^\dagger_{\ell m,\mathbf p}a_{1,\frac{\mathbf p}{2}+\mathbf k} a_{2,\frac{\mathbf p}{2}-\mathbf k}
+h.c.]\label{Lfd},
\end{align} 
where $V$ is the volume of the system. Since we focus on the effects of the $d$-wave resonance, we neglect possible background scatterings of either $s$- or $p$-wave symmetry, and those due to direct couplings between the fermions. The term proportional to $\eta=\pm 1$ describes the energy of a single dimer, with $\bar\nu$ being its detuning.  Unlike the case for $p$-wave scattering, an extra term proportional to the bare coupling constant $\bar z$ is constructed in order to renormalize the effective range $R$ [see Eq.~(\ref{r})], while still respecting the Galilean invariance. As will be shown later, it is necessary to take $\eta=-1$ in order to achieve a renormalizable theory.

The effective field theory in Eq.~(\ref{L}) differs from that for the $s$-wave and $p$-wave resonance models, and it is worthwhile to point out the differences. In the $s$-wave case, Kaplan was the first to use an $s$-wave dimer field $b_{00,\mathbf k}$ to describe the non-relativistic scattering between nucleons with a large $s$-wave ($\ell=0$) scattering length $a_s$~\cite{Kaplan1997}. In this case, the zero-range limit $\Lambda\to\infty$ is well defined with the choice $\eta=1$ and $\bar{z}=0$ by matching the scattering matrix with the $s$-wave phase shift expansion $k\cot\delta_s(k)=-1/a_s$. The same resonance model was constructed independently by Kokkelmans \emph{et. al.} for atoms close to an $s$-wave Feshbach resonance~\cite{Kokkelmans2002}, for which the dimer field $b_{00,\mathbf k}$ naturally represents the closed channel molecules.

Different from the $s$-wave case, low-energy scattering in the $p$-wave channel is described by two parameters, $k^3\cot\delta_p(k)=-1/v_p-k^2/R_p$~\cite{Braaten2012p}. Here $v_p$ is the $p$-wave scattering volume and $R_p$ is the $p$-wave effective range. In this case, however, to obtain a renormalizable theory with finite $v_p$ and $R_p$ in the low energy limit, one has to take $\eta=-1$. This means that the free dimer field $b_{1m,\mathbf k}$ becomes {\em ghost} field with negative norm~\cite{Braaten2012p}. However, such negative norm is only relevant at a much higher energy, of order of $\Lambda^2$, which is irrelevant for the low-energy physics described by the scattering phase shift $\delta_p(k)$. 
 
In the $d$-wave interaction resonance, it is first important to note that the low-energy scattering phase shift must be retained up to order $k^4$, namely $k^5\cot\delta_d(k)=-1/D-k^2/v-k^4/R$; the three interaction parameters $D$, $v$ and $R$ are the minimal set. 
This is because across the resonance, while the magnitude of $D$ can be tuned to be much larger than the interaction range $r_0$, $v/r_0^3$ and $R/r_0$ are typically of order unity. Taking the zero limit $v\to0$ or (and) $R\to0$ would lead to the noninteracting limit, i.e., $\delta(k)\to0$, which cannot describe the original interacting system. In contrast, it is safe to take the zero limit of the expansion coefficients of order higher than $k^4$ in $k^5\cot\delta(k)$. Now, we note that in Eq.~(\ref{L}), the term $d_{\ell m}^{\dagger}(i\partial_{t})d_{\ell m}$ corresponds to the total energy of two scattering fermions, and the term $d_{\ell m}^{\dagger}(-\nabla^2/4M)d_{\ell m}$ corresponds to the center of mass energy. The combination $d_{\ell m}^{\dagger}(i\partial_{t}+\nabla^2/4M)d_{\ell m}$ thus corresponds to the relative scattering energy. As a result, we explicitly construct the extra term $\bar{z} d_{\ell m}^{\dagger}(i\partial_{t}+\nabla^2/4M)^2d_{\ell m}$ in Eq.~(\ref{L}) to match the $k^4$-dependence of $k^5\cot\delta_d(k)$ for $d$-wave resonances. Note that by construction, the Lagrangian Eq.~(\ref{L}) maintains explicitly the Galilean invariance.

\begin{figure}[t]
\includegraphics[width=3 in]{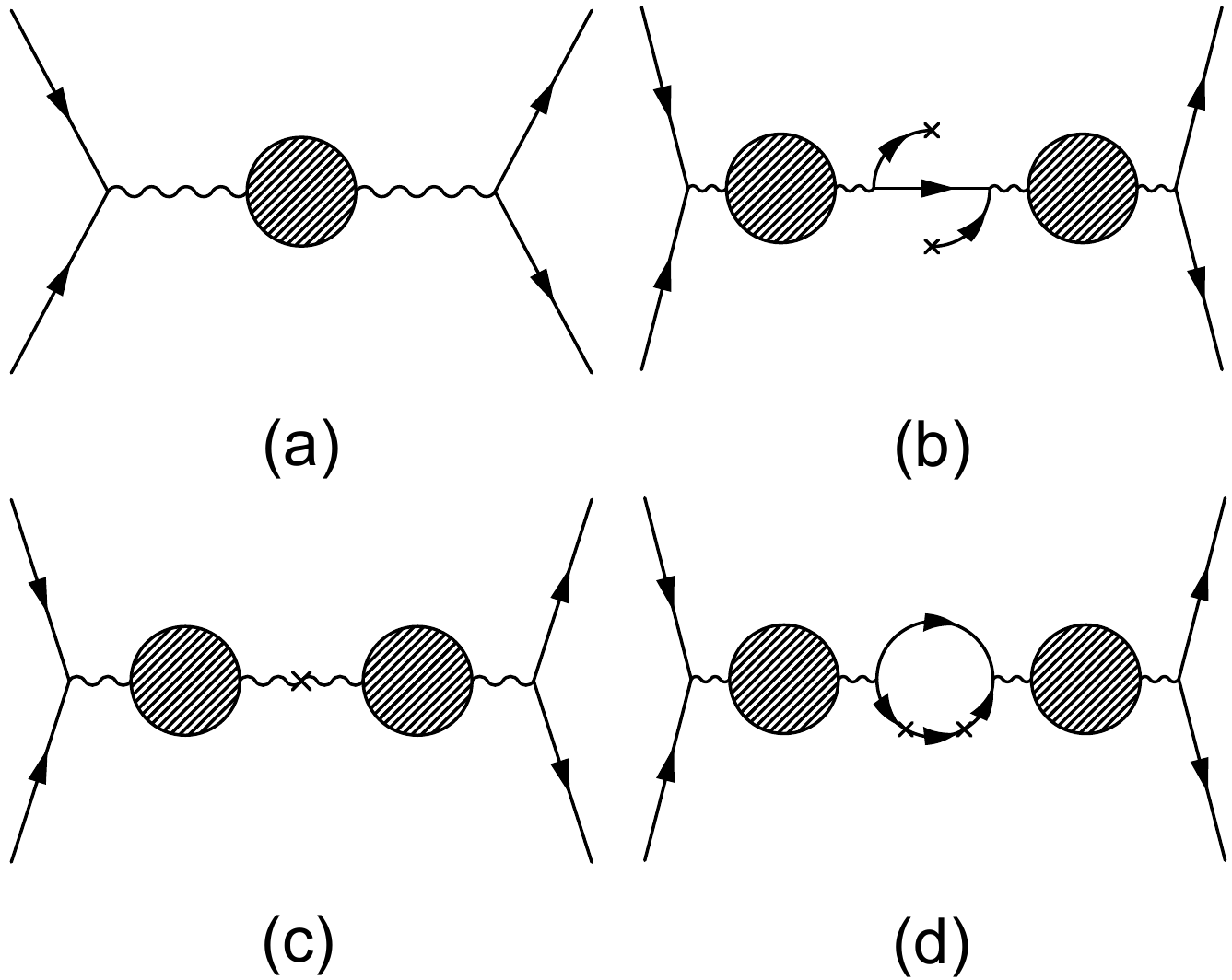}
\caption{Feynman diagrams for: (a) the $T$-matrix for two fermions; (b) the matrix element of $\psi^{\dagger}_i(\mathbf{R}+\mathbf{r}/{2})\psi_i(\mathbf{R}-\mathbf{r}/{2})$ ; (c) the matrix element of dimer bilinears; (d) the diagram for the Raman spectrum. In these diagrams, the wavy lines represent the propagators for the bare dimer fields, the solid lines represent the propagators for the bare fermion fields and the crosses represent the operators which are inserted.}
\label{dia1}
\end{figure}

The renormalizability of Eq.~(\ref{L}) is manifested by calculating the $T$-matrix, $T({\bf P},{\bf k},{\bf k}',\Omega)$, of scattering between two fermions with relative incoming (outgoing) momentum $2{\bf k}$ ($2{\bf k}'$) and total momentum ${\bf P}$. Due to the Galilean invariance of Eq.~(\ref{L}), one only needs to calculate in the center of mass frame, and the $T$-matrix is given by
\begin{equation}
T_m({\bf 0}, {\mathbf{k}},{\mathbf{k'}}, \Omega)=-4\pi\bar g^2 k^4Y_{2m}(\hat{\bf k})Y^*_{2m}(\hat{\bf k}')\mathcal D({\bf 0},\Omega),
\end{equation}
where $|{\bf k}|=|{\bf k}'|$ due to energy conservation and $\hat{\bf k}={\bf k}/|{\bf k}|$ and $\hat{\bf k}'={\bf k'}/|{\bf k'}|$. $\mathcal{D}({\bf P},\Omega)$ is the full dimer propagator, given in Fig.~\ref{dia1}(a)
\begin{align}
&\mathcal D^{-1}({\bf P},\Omega)\nonumber\\
&=\bar{\mathcal{D}}^{-1}({\bf P},\Omega)-\frac{\bar g^2}{2\pi^2}\int_0^\Lambda dq\frac{q^6}{\Omega-P^2/4M-q^2/M},
\end{align}
where $\bar{\mathcal{D}}({\bf P},\Omega)$ is the bare dimer propagator given by
\begin{align}
\bar{\mathcal{D}}(P,\Omega)=\frac{E_{p,+}-E_{p,-}}{\eta\bar z}\left(\frac1{\Omega-E_{p,+}}-\frac1{\Omega-E_{p,-}}\right),
\end{align}
with $E_{p,\pm}=P^2/4M-(1\mp\sqrt{1+4\bar\nu\bar z})/2\bar z$ the dimers' normal mode energies. In the case $1+4\bar\nu\bar z>0$, there always exits one branch of $\bar{\mathcal{D}}(P,\Omega)$ with negative weight corresponding to the presence of \emph{ghost} fields~\cite{Braaten2012}, {\em irrespective} of the sign of $\eta$. The appearance of ghost fields is inevitable due to the requirement to renormalize not only $v$ but also $R$ for $d$-wave interactions [see Eqs.~(\ref{v}) and (\ref{r})]~\cite{Braaten2012p}. 
In the case $1+4\bar\nu\bar z<0$, the poles of $\bar{\mathcal{D}}(P,\Omega)$ move away from the real axis into the complex plane and by itself seems problematic. However, the low energy observables predicted by the full coupled effective field theory remains valid (see below). In Table \ref{tab}, we summarize the main differences between our $d$-wave effective field theory with the 
$s$- and $p$-wave cases.

\begin{table}[h]
\begin{tabular}{m{1.2cm}  m{0.8 cm} m{2 cm} m{0.8 cm} m{0.8 cm} m{1.8 cm}} 
 \hline\hline
  & $\ell$ &  minimal parameters & $\eta$ & $\bar{z}$  & {\rm ghost field}\\ 
 \hline
 {\rm $s$-wave}  & $0$ &  $a_s$ & $1$ & 0 & No \\ 
 {\rm $p$-wave}  & $1$ &  $v_p, R_p$ & $-1$ & 0 & Yes \\ 
 {\rm $d$-wave}  & $2$ &  $D, v, R$ & $-1$ & $\neq 0$ & Yes \\ 
  \hline\hline
\end{tabular}
 \caption{Differences between our $d$-wave effective field theory with the 
$s$- and $p$-wave cases. Each cases are renormalized to the minimal interaction parameters listed.}
 \label{tab}
\end{table}

Matching $T_m({\bf 0}, k\hat{{\bf k}},k\hat{{\bf k}}', k^2/M+i0)$ with $\cot\delta_d(k)=-1/Dk^5-1/vk^3-1/Rk$ in the limit $k\to 0$, we find the renormalization conditions:
\begin{align}
\frac1{D} &=-\eta\frac{4\pi \bar\nu}{\bar g^2 M}+\frac{2\Lambda^5}{5\pi},\label{d}\\
\frac1{v} &=\eta\frac{4\pi }{\bar g^2 M^2}+\frac{2\Lambda^3}{3\pi},\label{v}\\
\frac1{R} &=\eta\frac{4\pi \bar z}{\bar g^2 M^3}+\frac{2\Lambda}\pi.\label{r}
\end{align}
To keep values of $D$, $v$ and $R$ finite while taking the limit $\Lambda\to\infty$, we require $\eta=-1$. Otherwise if $\eta=1$, from Eq.~(\ref{v}), $|v|<3\pi/2\Lambda^3$ and approaches zero. 
In fact, it turns out not possible to construct a purely fermionic model with contact inter-fermion interactions which reproduces the correct 
$d$-wave low energy scattering amplitude with finite parameters $v$ and $R$ in the limit $\Lambda\to\infty$. Thus it is crucial to introduce the dimer field with the concomitant appearance of the ghost field which, however, does not alter the low energy physics.

The applicable regime of our effective field theory can be analysed from the pole structure of $T_m$ in terms of the renormalized parameters
\begin{align}
&T_m({\bf 0}, k{\mathbf{\hat k}},k{\mathbf{\hat k'}}, \Omega)\nonumber\\
&=-\frac{16\pi^2 k^4Y_{2m}(\mathbf{\hat k})Y^*_{2m}(\mathbf{\hat k'})/M}{1/D+M\Omega/v+(M\Omega)^2/R+i(M\Omega)^{5/2}}.\label{rt}
\end{align}
For simplicity, let us consider the limit $1/D\to 0^+$. The real pole of $T_m$ at $\Omega\to0^-$ with positive weight $\sim v$ corresponds to a physical two-fermion bound state approaching threshold. However, since typically $v\sim r_0^3$ and $R\sim r_0$, 
there are other complex poles at energies $|\Omega|\sim1/Mr_0^2$, which apparently violate the unitary condition on the $S$-matrix. The origin of these unphysical poles is the truncation of $\cot\delta_d(k)$. However, as long as we are only interested in energy scales much smaller than $1/Mr_0^2$ as we shall do in the following, our effective field theory should give physically valid results.

\emph{$D$-wave contacts.} Effective field theory has served as an ideal formalism to elucidate the universal aspects of quantum gases~\cite{Nishida2006, Braaten2008b}; in particular, the derivation of universal relations involving the so-called contacts using the operator product expansion (OPE)~\cite{Braaten2008a, Braaten2010, Wilson, Braaten2008b, Braaten2008c, Son2010, Braaten2011, Hofmann2011, Zwerger2011, Braaten2012, Goldberger2012}. This is an operator relation for the product of two operators at small separation~\cite{Wilson, Peskin}
 \begin{align}
{O}_{i}\left(\mathbf{R}+\dfrac{\mathbf{r}}{2}\right){O}_{j}\left(\mathbf{R}-\dfrac{\mathbf{r}}{2}\right)
=\sum_{l} f^{ij}_{l}(\mathbf{r}){O}_{l}(\mathbf{R}) \label{ope}
 \end{align}
where $O_i$ are the {\em local} operators and $f^{ij}_l(\mathbf r)$ are the expansion functions. A similar expansion can also be carried out in the time domain. OPE is an ideal tool to explore short-range physics, $r_0\ll r\ll n^{-1/3}$ in a field theory context. Here $n$ is the average density.

In the case of $d$-wave interactions, we first define three contact densities (operators) as the derivatives of the Lagrangian density $\mathcal L$ with respect to $D^{-1}$, $v^{-1}$ and $R^{-1}$, by using Eqs.~(\ref{d}) to (\ref{r})
\begin{align}
\frac{\hat{\mathcal C}_D}{M} &\equiv\frac{\delta\mathcal L}{\delta (D^{-1})}=\frac{M\bar g^2}{4\pi} \sum_m d^\dagger_{\ell m} d_{\ell m},\label{cd}\\
\frac{\hat{\mathcal C}_v}{M} &\equiv \frac{\delta\mathcal L}{\delta (v^{-1})}
=\frac{M^2\bar g^2}{4\pi}\sum_m d^\dagger_{\ell m}\left(i\partial_t+\frac{\nabla^2}{4M}\right) d_{\ell m},\label{cv}\\
\frac{\hat{\mathcal C}_R}{M} &\equiv \frac{\delta\mathcal L}{\delta (R^{-1})}=\frac{M^3\bar g^2}{4\pi}\sum_m d^\dagger_{\ell m}\left(i\partial_t+\frac{\nabla^2}{4M}\right)^2 d_{\ell m}.\label{cr}
\end{align}
Note that we have used the equation of motion satisfied by $d_{\ell m}$ to obtain the concise expression of Eq.~(\ref{cv}). While $\hat{\mathcal C}_D$ is proportional to the total dimer density, $\hat{\mathcal C}_v$ and $\hat{\mathcal C}_R$ can be considered as proportional to the ones weighted by the powers of the internal energy of the dimers. A similar structure has been found for $p$-wave contacts~\cite{Yu2015}. In addition, as we will see from the tails of the momentum distribution and the Raman spectroscopy, it is also useful to introduce two extra $d$-wave contact densities as
\begin{align}
\frac{\hat{\mathcal C}_{D,P}}{M} &\equiv\frac{M^2\bar g^2}{4\pi}\sum_m d^\dagger_{\ell m}\left(-\frac{\nabla^2}{4M}\right) d_{\ell m}\label{cdp},\\
\frac{\hat{\mathcal C}_{v,P}}{M} &\equiv\frac{M^3\bar g^2}{4\pi}\sum_m d^\dagger_{\ell m}\left(i\partial_t+\frac{\nabla^2}{4M}\right)\left(-\frac{\nabla^2}{4M}\right) d_{\ell m}\label{cvp},
\end{align}
which, compared with Eqs.~(\ref{cd}) and (\ref{cv}), are further weighted by the kinetic energy of the dimers, and encapsulate additional information of correlations at short distances. 
The spatial integration of the expectation values of the contact densities are defined as the $d$-wave contacts: $C_D=\int d\mathbf r \langle\hat{\mathcal C}_D\rangle$, $C_v=\int d\mathbf r \langle\hat{\mathcal C}_v\rangle$, $C_R=\int d\mathbf r \langle\hat{\mathcal C}_R\rangle$, $C_{D,P}=\int d\mathbf r\langle\hat{\mathcal C}_{D,P}\rangle$, and $C_{v,P}=\int d\mathbf r \langle\hat{\mathcal C}_{v,P}\rangle$. From Eqs.~(\ref{cd}-\ref{cr}), one can write down the adiabatic theorems,
\begin{equation}
\frac{\partial F}{\partial \alpha^{-1}}=-\frac{C_\alpha}{M};~~\alpha=D, v, R,\label{df}
\end{equation}
where $F$ is the free energy of the system. To illustrate the use of the effective field theory, we now derive some universal relations between the introduced contacts and various physical observables.

\emph{Short distance expansion.} 
The tails of the momentum distribution can be extracted from the one-body density matrix $\rho_i(\mathbf R,\mathbf r)=\langle\psi_i^{\dagger}(\mathbf{R}+\mathbf{r}/2)\psi_i(\mathbf{R}-\mathbf{r}/2)\rangle$ and can be measured experimentally by the time-of-flight technique \cite{Jin2010, Jin2014}. To relate $\rho_i(\mathbf R,\mathbf r)$ with the $d$-wave contacts, we calculate the OPE by matching the matrix elements of operators from an incoming state $|I\rangle$ with two fermions of different species having momentum $\mathbf P/2+k \hat{\mathbf{k}}$ and $\mathbf P/2-k \hat{\mathbf{k}}$  to an outgoing state $|F\rangle$ with two fermions having momentum $\mathbf P/2+k \hat{\mathbf{k}}'$ and $\mathbf P/2-k \hat{\mathbf{k}}'$. The total energy of the fermion pair is $E=P^2/4M+k^2/M$. Since we are interested in the rotationally invariant case, we will average over the direction of the total momentum $\mathbf P$. The case without rotational invariance can be calculated similarly. The matrix element of $\rho_i$ is given by the diagram shown in Fig.~\ref{dia1}(b) and 
the result is
\begin{align}
& \langle F|\rho_i(\mathbf R,\mathbf r)|I\rangle=4\pi M^2\bar g^{4}k^4 \sum_{m}Y_{2m}(\hat{\mathbf k})Y^{*}_{2m}(\hat{\mathbf k}')\mathcal D^{2}(P,E)\nonumber\\
&\times\left[\delta(\mathbf{r})+\dfrac{k^2}{2\pi r}-\dfrac{3r(k^4+P^2k^2/18)}{8\pi}\right]+{\rm const.}+o(\mathbf r).\label{rho}
\end{align}
Likewise, we calculate the matrix elements of the contact densities according to the diagrams shown in Fig.~\ref{dia1}(a, b).  
We find
\begin{align}
\langle F|\hat{\mathcal C}_D|I\rangle &=M^2\bar g^{4}k^4\sum_m Y_{2m}(\hat{\mathbf k})Y^{*}_{2m}(\hat{\mathbf k}')\mathcal D^{2}(P,E),\label{ncd}\\
\langle F|\hat{\mathcal C}_v|I\rangle &=k^2\langle F|\hat{\mathcal C}_D|I\rangle,\label{ncv}\\
\langle F|\hat{\mathcal C}_R|I\rangle &=k^4\langle F|\hat{\mathcal C}_D|I\rangle,\label{ncr}\\
\langle F|\hat{\mathcal C}_{D,P}|I\rangle &=P^2\langle F|\hat{\mathcal C}_D|I\rangle/4,\label{ncdk}\\
\langle F|\hat{\mathcal C}_{v,P}|I\rangle &=P^2k^2\langle F|\hat{\mathcal C}_D|I\rangle/4.\label{ncvk}
\end{align}

After Fourier transforming Eq.~(\ref{rho}) and matching with Eqs.~(\ref{ncd}) to (\ref{ncvk}), we find that the momentum distribution $n_i(\mathbf q)$ of the $i$th species has a tail in the large $q$-limit ($n^{1/3}\ll q\ll 1/r_0$)
\begin{align}
n_{i}(\mathbf q)=\frac{1}{V}\left[\dfrac{C_{D}}{2\pi^{2}}+\dfrac{C_{v}}{\pi^{2}q^{2}}+\dfrac{9C_{R}+2C_{v,P}}{6\pi^{2}q^{4}}\right],\label{tail}
\end{align} 
whose magnitude depends on the $d$-wave contact densities. The presence of the additional quantity $C_{v,P}$, which can not be derived from the adiabatic theorems (\ref{df}), in the momentum tail can be understood in the following way. Let us consider a single pair of interacting fermions. In the center of mass frame of the pair where $C_{v,P}$ is zero according to Eqs.~(\ref{cvp}) and (\ref{tail}), the momentum tail $n_{\rm com}(\mathbf q)$ involves only $C_\alpha$ for $\alpha=D,v,R$. However, when we switch to a reference frame moving with a relative velocity $\mathbf u$, the momentum tail of the pair in this new frame should be $n(\mathbf q)=n_{\rm com}(\mathbf q-m\mathbf u)$. Expansion of $n_{\rm com}(\mathbf q-m\mathbf u)$ to order $1/q^4$ leads to 
an extra term $\sim u^2 C_v$ in $n(\mathbf q)$, which is exactly the generally nonzero $C_{v,P}$ term in Eq.~(\ref{tail}) in this case.
Note that the Galilean invariance garrauntees $C_D$ and $C_v$ having the same values in different reference frames [cf.~Eq.~(\ref{df})]. Quantities similar to $C_{v,P}$ have been introduced for $p$-wave interactions in three dimensions~\cite{Yu2015, Peng2016x, Yi2016x}. 

The tails of the momentum distribution $n_{i}(\mathbf q)$ seems to yield a divergent number of fermions. Actually, by the $U(1)$ gauge invariance of Eq.~(\ref{L}), the conserved total particle number is given by
\begin{align}
\hat{N}=&\int d\mathbf r\Big\{\sum _{i=1,2}\psi_{i}^{\dagger}\psi_{i}\nonumber\\
-&\sum_{m}(d_{m}^{\dagger}\left[1+\bar z \left(2i\partial_{t}+{\nabla^2}/{2M}\right)\right]d_{m}+h.c.)\Big\}.
\end{align}
Using the renormalization relations (\ref{d}), (\ref{v}) and (\ref{r}), one can verify that the divergent part of $n_{i}(\mathbf q)$ at large $q$ is cancelled by the dimer terms; the dimer terms can be considered as counterterms to the fermion densities. Note that the factor $\bar z \left(2i\partial_{t}+{\nabla^2}/{2M}\right)$ is due to the expansion of the bare dimer fields in terms of their normal modes.

\emph{Short distance and time expansion.} Single-particle spectral function, which reveals fundamental properties of an interacting many-body system, such as pairing and pseudo-gap phenomena, can be measured using Raman spectroscopy in atomic gases~\cite{Gaebler2010,Feld2011}. When two Raman lasers of frequency $\omega_1$ and $\omega_2$ and wave-vector $\mathbf k_1$ and $\mathbf k_2$ are applied, atoms can be excited from the initial internal state $|2\rangle$ to the final internal state $|3\rangle$ by absorbing energy $\omega=|\omega_1-\omega_2|$ and momentum $\mathbf q=\mathbf k_1-\mathbf k_2$. The resultant number of atoms transferred to state $|3\rangle$ is, by the Fermi golden rule, proportional to the rate
\begin{align}
I_{\rm Ra}(\mathbf{q},\omega)=&-\dfrac{1}{\pi}{\rm Im}\Pi_{\rm Ra}(\mathbf{q},\omega), \\
 \Pi_{\rm Ra}(\mathbf{q},\omega)=&-iV\int dtd\mathbf{r} \,e^{i\omega t-i\mathbf{q}\cdot\mathbf{r}}\langle T \mathcal Q_{23}(\mathbf{r},t)\mathcal Q_{23}^{\dagger}(\mathbf{0},0)\rangle,
\end{align}
with $\mathcal Q_{23}(\mathbf{r},t)\equiv \psi^{\dagger}_{3}(\mathbf{r},t)\psi_{2}(\mathbf{r},t)$.

By calculating the OPE of $\mathcal Q_{23}(\mathbf{r},t)\mathcal Q_{23}^{\dagger}(\mathbf{0},0)$ in both the time and space domain, we find for $\omega>\epsilon_q\equiv q^2/2M$: 
\begin{align}
&\frac\pi M I_{\rm Ra}(\mathbf{q},\omega)
={\left(M\omega-\dfrac{q^2}{4}\right)^{1/2}{C}_{D}}-\dfrac{q^2C_{D,P}}{3(4M\omega-q^2)^{3/2}}\nonumber \\
&+\left[\dfrac{q}{\sqrt{4M\omega-q^2}}+4\sinh^{-1}\left(\dfrac{q}{\sqrt{4M\omega-2q^2}}\right)\right]\frac{{C}_{v}}q\nonumber \\
&+\dfrac{2q^2(7q^4-40q^2M\omega+60M^2\omega^2)}{3(2M\omega-q^2)^2(4M\omega-q^2)^{5/2}}C_{v,P}\nonumber \\&+\dfrac{q^4-20q^2M\omega+60M^2\omega^2}{(2M\omega-q^2)^2(4M\omega-q^2)^{3/2}}{C}_{R}.\label{ra}
\end{align}
For $\epsilon_q>\omega>\epsilon_q/2$, $I_{\rm Ra}(\mathbf{q},\omega)$ is given by Eq.~(\ref{ra}) with the factor $\sinh^{-1}[q/\sqrt{4M\omega-2q^2}]$ replaced by $\cosh^{-1}[q/\sqrt{-4M\omega+2q^2}]$. $I_{\rm Ra}(\mathbf{q},\omega)=0$ when $\omega<\epsilon_q/2$. In the limit $q\to 0$, $I_{\rm Ra}(\mathbf{0},\omega)$ gives the radio-frequency response and involves only $C_v, C_D$ and $C_R$. The presence of $C_{D,P}$ and $C_{v,P}$ in Eq.~(\ref{ra}) can also be understood from a Galilean covariance argument similar to the one given below Eq.~(\ref{tail}).

{\em Discussion}. The construction of the effective field theory Eq.~(\ref{L}) for $d$-wave resonance suggests a general procedure for resonances of arbitrary higher partial waves. Consider a two-component Fermi gas with short-range interactions, the phase shift in the $\ell$-th scattering channel can be written as $k^{2\ell+1}\cot\delta_\ell(k)=-\sum_{\alpha=0}^{\ell} k^{2\alpha}/a_{\ell\alpha}+O(k^{2\ell+2})$ in the low energy limit. To reproduce the phase shift, we need only to generalize the dimer field term in Eq.(\ref{L}) to
\begin{equation}
\mathcal{L}_d=\sum_{m=-\ell}^{\ell} \sum_{\alpha=0}^{\ell}d_{\ell m}^{\dagger}\bar z_{\ell \alpha}\left(i\partial_{t}+\dfrac{\nabla^{2}}{4M}\right)^{\alpha}d_{\ell m},
\end{equation}
and assume $L_{fd}$ to be the form of Eq.~(\ref{Lfd}) with the factor $\bar g\sqrt{4\pi/V}$ replaced by $4\pi/\sqrt{MV}$, which amounts to a rescaling of the dimer field $d_{\ell m}$.
The relation between parameters $\{\bar z_{\ell \alpha}\}$ to the physical scattering parameters $\{a_{\ell\alpha}\}$ can be established similarly by matching the scattering $T$-matrix to that of $k^{2\ell+1}\cot\delta_\ell(k)$. One finds
\begin{align}
\frac1{a_{\ell\alpha}}=\bar z_{\ell\alpha}M^\alpha+\frac2{\pi}\frac{\Lambda^{2(\ell-\alpha)+1}}{2(\ell-\alpha)+1},
\end{align}
for $0\le\alpha\le\ell$.
For fixed $\{a_{\ell\alpha}\}$, the zero range limit $\Lambda\to\infty$ is attainable only if $\bar z_{\ell\alpha}$ are all negative. Our formalism sets the stage for the exploration of universal aspects of both few-body and many-body physics close to a higher partial wave resonance. Further important questions remain to be investigated, including the effects of long-range and multi-body interactions.

\textit{Acknowledgment.} We thank Hui Zhai, Ling-Fong Li, Zheyu Shi and Yusuke Nishida for helpful discussions. This work is supported by Tsinghua University Initiative Scientific Research Program, NSFC Grant No.~11474179. SZ is supported by Hong Kong Research Grants Council (General Research Fund, HKU 17306414 and Collaborative Research Fund, HKUST3/CRF/13G) and the Croucher Innovation Awards.

\end{document}